\documentclass{iau} 
\usepackage{graphicx}
\usepackage{amsmath,amsfonts,amssymb}

\newcommand{\n}[0]{\vec{n}}

\title[Bayesian component separation: The \emph{Planck} experience] %% give here short title %%
{Bayesian component separation:\\ The \emph{Planck} experience}

\author[Wehus \& Eriksen]   %% give here short author list %%
{Ingunn Kathrine Wehus$^1$ and Hans Kristian Eriksen$^1$}

\affiliation{$^1$Institute of Theoretical Astrophysics, University of Oslo, \\ Postboks 1029 Blindern,
  0315 Oslo, Norway \\ email: {\tt i.k.wehus@astro.uio.no}}

\pubyear{2018}
\volume{333}  %% insert here IAU Symposium No.
\jname{Peering towards Cosmic Dawn}
\editors{Vibor Jeli\'c \& Thijs van der Hulst, eds.}
\begin{document}

\maketitle

\begin{abstract}
  Bayesian component separation techniques have played a central role
  in the data reduction process of \emph{Planck}. The most important
  strength of this approach is its global nature, in which a
  parametric and physical model is fitted to the data. Such physical
  modeling allows the user to constrain very general data models, and
  jointly probe cosmological, astrophysical and instrumental
  parameters. This approach also supports statistically robust
  goodness-of-fit tests in terms of data-minus-model residual maps,
  which are essential for identifying residual systematic effects in
  the data. The main challenges are high code complexity and
  computational cost. Whether or not these costs are justified for a
  given experiment depends on its final uncertainty budget. We
  therefore predict that the importance of Bayesian component
  separation techniques is likely to increase with time for intensity
  mapping experiments, similar to what has happened in the CMB field,
  as observational techniques mature, and their overall sensitivity
  improves.  \keywords{Cosmology: cosmic microwave background, Galaxy:
    general, Methods: data analysis}
\end{abstract}

\firstsection % if your document starts with a section,
\section{Introduction}

On May 14th 2009, ESA's cosmology cornerstone satellite mission
\emph{Planck} (\cite[Planck Collaboration I 2016]{planck_mission})
launched from French Guyana. For the next four years, the satellite
made the deepest full-sky measurements of the microwave sky to date,
observing in nine frequencies between 30 and 857~GHz. Based on these
measurements, the \emph{Planck} collaboration was able to constrain
the parameters of our currently best $\Lambda$CDM cosmological model
to unprecedented accuracy (\cite[Planck Collaboration XIII
  2016]{planck_parameters}); measure the angular power spectrum of
weak gravitational lensing with $40\sigma$ significance (\cite[Planck
  Collaboration XV 2016]{planck_lensing}); rule out all
inflationary models predicting a high degree of non-Gaussianity
(\cite[Planck Collaboration XVII 2016]{planck_fnl}); and open up a
whole new view of the various emission mechanisms in the Milky Way
(\cite[Planck Collaboration X 2016]{planck_foregrounds}), just to mention a few
highlights. Today, \emph{Planck} represents a standard reference in
the field of cosmology, and is likely to remain so in the foreseeable
future.

\begin{figure}[t]
% \vspace*{-2.0 cm}
\begin{center}
  \includegraphics[width=0.7\columnwidth]{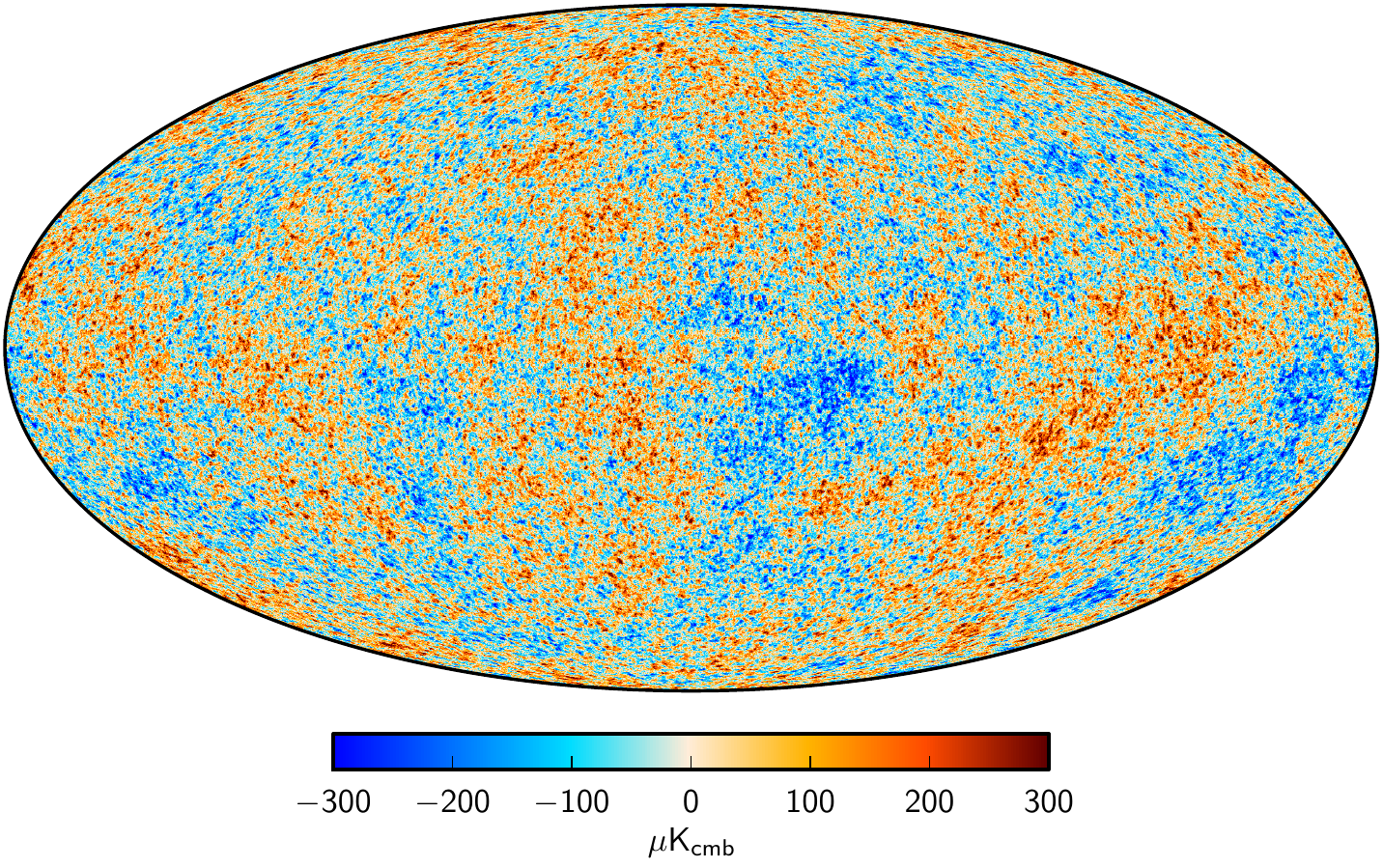} \\
  \includegraphics[width=0.7\columnwidth]{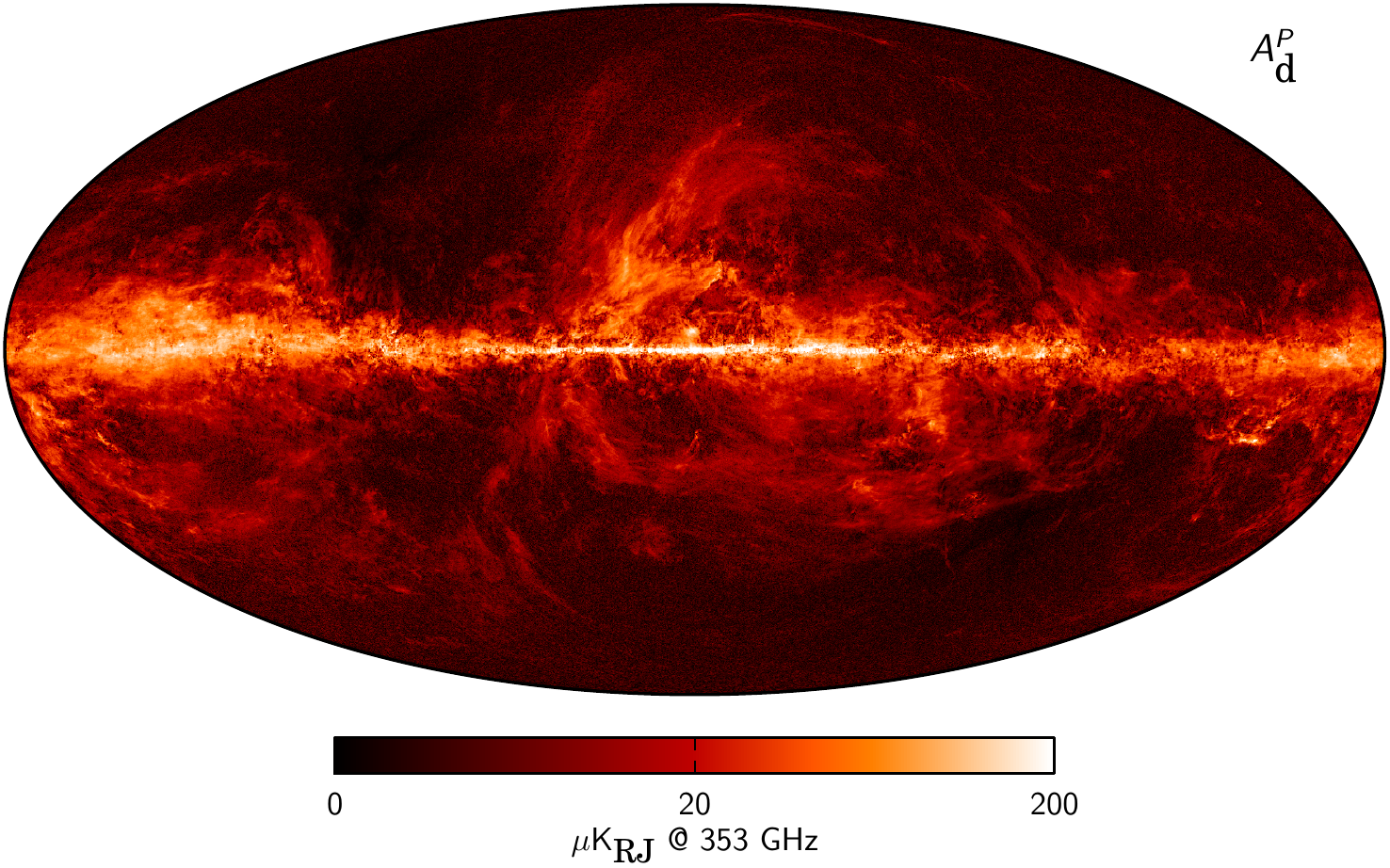} 
% \vspace*{-1.0 cm}
 \caption{(\emph{Top:}) The \emph{Planck} 2015 Commander CMB
   temperature fluctuation map at $5'$ FWHM angular
   resolution. (\emph{Bottom:}) The \emph{Planck} 2015 Commander
   thermal dust polarization amplitude map at $10'$ FWHM
   resolution, evaluated at 353\,GHz. Both plots are reproduced from
   \cite{planck_mission}.}
   \label{fig1}
\end{center}
\end{figure}

These achievements are the direct results of an intimate collaboration
between hundreds of engineers and scientists spanning more than 25
years, resulting in a number of breakthroughs in both instrumentation
and data analysis techniques. In this short note, we will discuss one
specific example of such breakthroughs, namely that of \emph{Bayesian
  component separation} (\cite[Eriksen et al. 2008]{eriksen2008};
\cite[Planck Collaboration X 2016]{planck_foregrounds}), on which we
have been working for the last five years. The primary goal of this
line of work is to establish the cleanest possible representation of
the CMB sky signal based on some set of ``dirty'' frequency maps,
which includes both cosmological signal, astrophysical foregrounds
originating both from our own Milky Way and beyond, and instrumental
systematics and noise. Of course, this challenge is the topic of many
competing algorithms, and there is no time or space to review all of
these methods here (see, e.g., \cite[Leach et al. 2008]{leach2008}).

What sets the Bayesian approach apart from most other techniques,
though, is its insistence on defining a \emph{physically motivated}
data model, including both the signal of primary interest (for us, the
CMB) and any contaminating contributions, whether it is from
synchrotron or thermal dust emission, or from the instrument
itself. This explicit parametric model is then fitted directly to the
raw data in the posterior distribution sense using standard
statistical Monte Carlo methods, such as Metropolis-Hastings or Gibbs
sampling, or simply non-linear optimization. Thus, from a statistical
point of view this framework represents a very direct and simple
approach to component separation, as it almost exclusively uses
methods that are routinely taught in introductory statistics
courses. At the same time, the approach is also extremely bold, in
that it aims to define a statistical description of the entire data
set, and not only the primary signal of interest.

To make this point more explicit, let us consider the data model
adopted for the latest \emph{Planck} 2015 temperature analysis, as
presented in \cite[Planck Collaboration X
  (2016)]{planck_foregrounds},
\begin{align}
  d_{\nu}(\theta) & = g_{\nu} \sum_{i=1}^{N_{\textrm{comp}}}
  \mathcal{F}_\nu^i(\beta_i, \Delta_\nu) a_{i} + \mathcal{T}_{\nu}
  m_{\nu} + n_{\nu}.
\end{align}
Here $d_{\nu}$ denotes the observed data at frequency $\nu$; the sum
runs over all relevant astrophysical components (e.g., CMB,
synchrotron, free-free, spinning and thermal dust, and CO and other
line emission mechanisms for the analysis in question), each of which
is parameterized by some mixing matrix $\mathcal{F}_\nu^i$ with
spectral parameters $\beta_{i}$, and an amplitude $a_{i}$ per pixel;
$g_{\nu}$ and $\Delta_{\nu}$ represent instrumental parameters
describing overall calibration and bandpass uncertainties,
respectively; $\mathcal{T}_{\nu}$ indicates fixed templates with free
amplitudes only, and accounts for zero-level and dipole uncertainties;
and $\n_{\nu}$ denotes instrumental noise. In total, this model
includes more than 300 million free and strongly correlated
parameters, and jointly constraining these represents a major
computational challenge. To this aim, we have over the last decade
developed a computer code called \emph{Commander} (\cite[Eriksen et
  al. 2004]{eriksen2004}, \cite[2008]{eriksen2008}), which implements
a particular MCMC sampling algorithm called Gibbs sampling that maps
out the exact, full and complicated posterior distribution through
iterative sampling over much simpler conditional distributions. This
code has become a standard reference in the field, and many of the
official \emph{Planck} products have been derived with this code. Two
particularly well-known examples are shown in Figure~1, namely the CMB
temperature fluctuation map (\emph{top panel}) and the thermal dust
polarization amplitude map at 353~GHz (\emph{bottom panel}).

Based on the \emph{Planck} experience, there is no question that the
Bayesian approach is both powerful, flexible and highly
productive. However, it is clearly also associated with a significant
cost in terms of both manpower and computing resources, and future
experimentalists may therefore ask themselves whether it makes sense
to adopt an equally ambitious approach for their experiment. Giving
some inputs to this question is the main goal of the current note.

%\vspace*{-0.5cm}
\section{What has Bayes ever done for us?}

\begin{figure}[t]
% \vspace*{-2.0 cm}
\begin{center}
 \includegraphics[width=0.89\columnwidth]{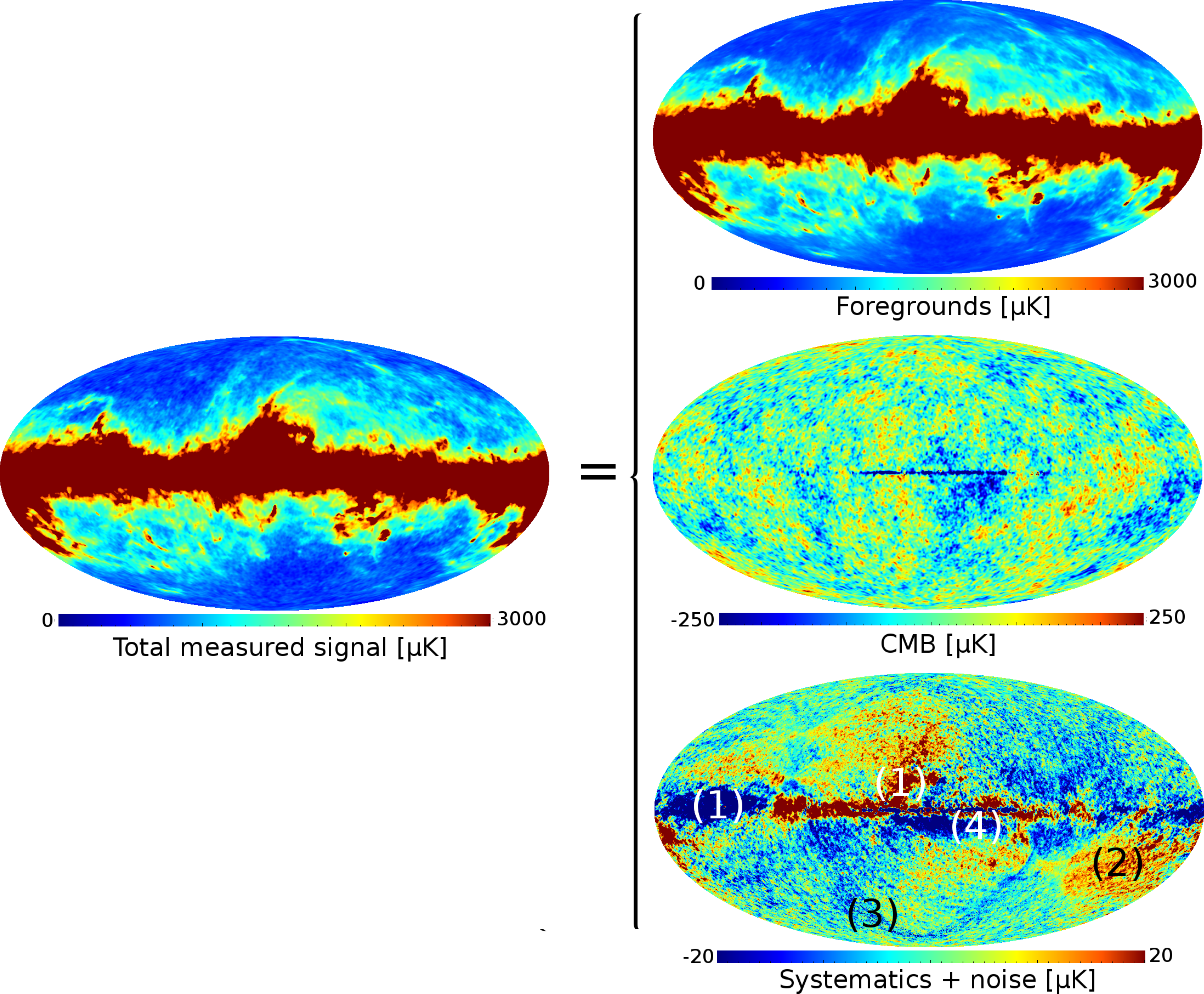} 
% \vspace*{-1.0 cm}
 \caption{Decomposition of the \emph{Planck} 2015 353-2 detector map
   (\emph{left panel}) into Galactic foregrounds (\emph{right top
     panel}), CMB fluctuations (\emph{right middle panel}), and
   residual systematics and noise (\emph{right bottom panel}). The
   latter three maps are all derived with Bayesian component
   separation, as implemented in Commander. The dominant instrumental
   systematic features seen in the bottom right panel correspond to
   (1) polarization-to-temperature leakage; (2) bandpass mismatch; (3)
   sidelobe pickup; and (4) transfer function mis-estimation. Note
   that none of these effects can be identified in the raw map, but
   they become visible only after component separation, when
   subtracting off a best-fit physical model.}
   \label{fig2}
\end{center}
\end{figure}

When the Bayesian approach to CMB component separation was first
introduced in 2004 by \cite{jewell2004}, \cite{wandelt2004} and
\cite{eriksen2004}, the main motivation was to establish a coherent
analysis framework that allowed for joint analysis of both CMB and
astrophysical foreground signals, and, critically, to propagate their
joint uncertainties from raw data products to final cosmological
parameters. After going through the \emph{Planck} experience, these
ideas do remain important, and play important roles in the final
results. However, even more important than these aspects for the
overall final outcome of \emph{Planck} is the insistence of
\emph{physical modeling} for all signal components.

For \emph{Planck}, this point has had two important consequences. On
the one hand, it implies that parameters that are normally considered
``nuisance'' parameters for cosmologists, such as thermal dust or CO
emission maps, actually become valuable scientific products in their
own rights. Furthermore, by virtue of being derived in precisely the
same manner as the CMB signal, they are ensured the same high
statistical confidence as the primary scientific target. As a result,
the Commander approach delivered not only a CMB fluctuation map, but
also individual estimates of synchrotron, free-free, spinning and
thermal dust, and CO line emission (\cite[Planck Collaboration X
  2016]{planck_foregrounds}), all of which are used extensively by
other groups.

On the other hand, there is in fact another aspect of the physical
modeling approach that is even more important than the derivation of
scientific ancillary products, and that is the method's unique ability
to uncover \emph{instrumental systematic effects}. To make this point
explicit, we show in the left panel the publicly released
\emph{Planck} 2015 353~GHz map derived from the 353-2 bolometer alone
(\cite[Planck Collaboration VIII 2016]{planck_hfi}). By visual
inspection, this map appears to be a very high signal-to-noise
measurement of the true sky at 353~GHz, and no obvious artifacts may
be seen. However, in the right column we also show the (sum of the)
astrophysical foregrounds at this channel (\emph{top panel}), as
estimated by Commander, and the corresponding CMB fluctuation map
(\emph{middle panel}). The bottom panel shows what we call the
residual map, which is nothing but the difference between the data and
the signal model, and encaptures unmodeled instrumental systematic
effects and noise. Ideally, this map should look like uncorrelated
noise with a local variance given by the observation time of the
instrument in a given pixel, but clearly it does not. Instead, a wide
range of residual instrumental systematic effects may be seen, and
understanding and mitigating the features in this map is really where
the vast majority of the analysis time for a given experiment is
spent.

When looking at a residual map like this, the important goal is to
understand each coherent feature physically; what causes it, and how
can it be fixed? For instance, in our case the Galactic plane
structures marked by (1) were shown to be correlated very strongly
with the thermal dust polarization signal shown in the bottom panel of
Figure~1, with a sign given by the detector orientation of the
\emph{Planck} detector. This effect could thus be traced down to
\emph{polarization-to-temperature leakage} through sub-optimal
polarization reconstruction. In collaboration with the map making
team, the issue could then be mitigated during low-level
processing. Likewise, the diffuse structures marked by (2) were traced
to bandpass differences between individual bolometers, leaking
foreground differences between bolometers into the noise component,
which then subsequently were spread around the sky by the noise
filter, resulting in what were internally in \emph{Planck} referred to
as ``dolphins''. This effect could then be improved by introducing
various bandpass corrections. Similarly, the effects marked by (3) and
(4) could be associated with sidelobe and transfer function
non-idealities, respectively.

We believe this simple example provides the single most clear
demonstration of the strength of the Bayesian approach to component
separation available today: By forcing oneself to work within a
physical and rigid model at the level of the raw data, in which no
arbitrary parameters or ``fudge factors'' are allowed, even subtle
non-idealities are revealed through statistically rigorous
goodness-of-fit tests and residual maps. Furthermore, these tests
provide far more intuitive information than methods that, say, live
purely in power spectrum domain, as it is much easier to visually
identify a given systematic effect in terms of residual map than a
power spectrum. As is often stated, ``seeing is believing'', and that
holds very true for high signal-to-noise data analysis.

%\vspace*{-0.7cm}
\section{Prospects for Bayesian analysis of intensity mapping experiments}

\begin{figure*}
    \centering
%    \raisebox{-0.5\height}{
%        \includegraphics[width=0.6\textwidth]{{comap_30-34ghz_2.5deg2_halos_3dvol_cropped}.pdf}
%    }
%    \\
    \raisebox{-0.5\height}{
        \includegraphics[width=0.8\textwidth]{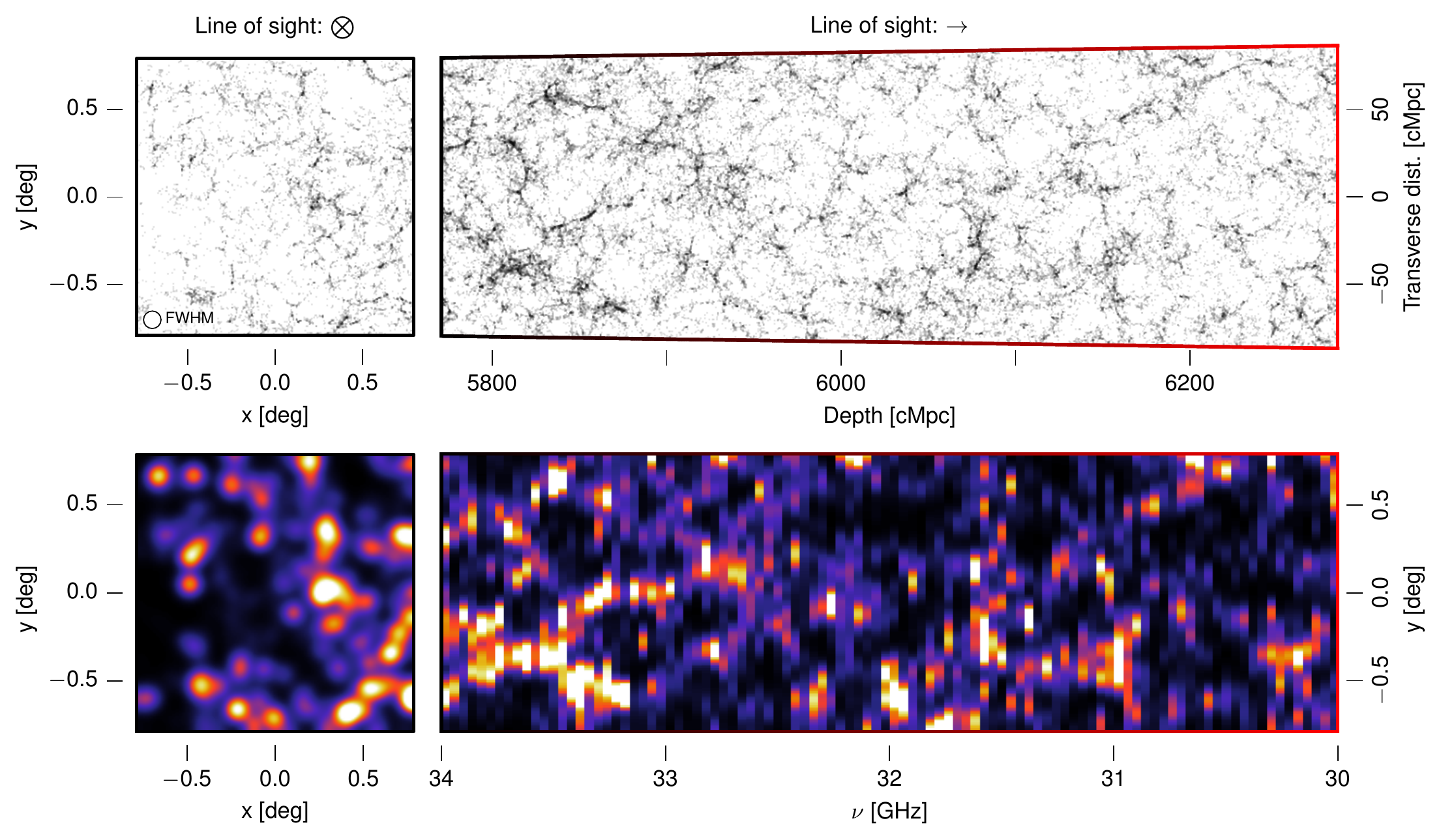}
    }
    \\
        \includegraphics[width=0.4\textwidth]{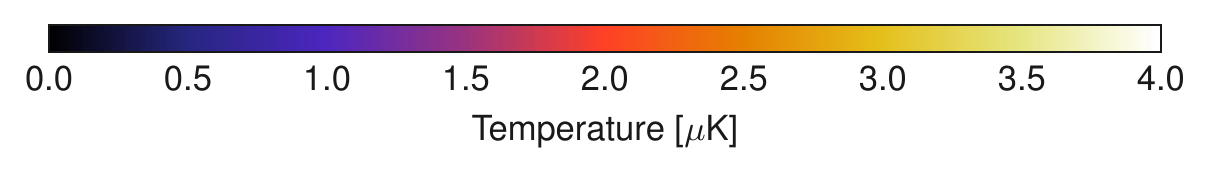}
    \caption{
        Simulated 3D volume as observed by the COMAP CO intensity
        mapping experiment; figure and caption are reproduced from \cite{li2016}.
%        (\emph{Top}:) Halos in the 3D volume, rendered to scale in
%        comoving distance.  The equivalent cosmological redshifts and
%        redshifted CO(1-0) frequencies are marked along the line-of-sight.
        (\emph{Top}:) 2D projections of halo positions.  The left
        image shows the ``front'' view of halos that would fall into
        the highest frequency channel, or lowest redshift
        slice.  The COMAP pathfinder beam size is shown for scale.  The right image shows the ``side'' view of halos to a depth of 6 arcmin, or one beam width.
        (\emph{Bottom}:) CO intensity map produced by the fiducial
        model of \cite{li2016}. The same large-scale structure is
        readily apparent in both images, even with the lower
        resolution of the intensity map. Note the colour scale
        ranging between 0 and $4\,\mu\textrm{K}$.
    }
    \label{fig:halos2imap}
\end{figure*}

As the cosmology community is gearing up towards mapping the full 3D
distribution of matter in the universe, the intensity mapping approach
has appeared as a particularly promising technology for surveying
large volumes in short time. Many of the observational techniques
required for this are similar to those used for CMB analysis, and it
is therefore appropriate to ask whether also the Bayesian approach
described above may turn out to be valuable for intensity mappers.

To answer this question, we believe that the \emph{Planck} experience
may provide some useful guidance. And the most important lesson
learned from that work is that the importance of full joint and
physical analysis increases with the signal-to-noise ratio of the data
set in question. When the data are largely dominated by ``white
noise'', the substantially higher computational and implementational
cost of a Bayesian approach is unlikely to be justified; it
\emph{does} take longer to write a full-blown Gibbs sampler than an
internal linear combination method, and the results are very
comparable for noise-dominated data. However, when the data eventually
mature, and the signal-to-noise ratio increases, more and more
systematic effects become visible above the noise floor, both in the
form of astrophysical and instrumental effects -- and in particular in
terms of interplay between the latter two. And it is in this regime
that the Bayesian approach is uniquely powerful.

We are personally involved in the CO intensity mapping experiment
called COMAP (\cite[Li et al. 2016]{li2016}), which will start
observations in early 2018. It will field 19 coherent detectors on a
10-meter telescope, each one coupled to a ROACH that provides a
spectral resolution of up to 4096 frequencies between 26 and
34~GHz. In total, this setup will cover redshifts between $z=2.4$ and
3.4. The first observations are, however, designed as a pathfinder, and the
expected random noise after 2 years of integration is
$\sim$10$\mu\mathrm{K}$ per resolution element. For comparison, the
bottom panel of Figure~3 shows the predicted CO signal for a slice
through the 3D volume, as simulated by \cite{li2016}; note that the
colour range spans from 0 to 4$\mu\mathrm{K}$. Thus, the
signal-to-noise ratio will be no larger than 0.4 per resolution
element. In this situation, a global Bayesian approach is unlikely to
yield any significant advantages over far simpler and computationally
cheap approaches.

However, in the second phase of the COMAP experiment, the plan is to
expand the array by a factor of six in the number of detectors, and
increase the observation time to at least three years, leading to
an increase in sensitivity per resolution element of at least
$\sqrt{1.5\cdot6}\,=\,$3. The signal-to-noise ratio
will then approach unity, and at that stage it becomes important to
properly account for subtle instrumental effects.

Based on our experience with \emph{Planck}, we are in COMAP currently
working toward a simple and computationally fast analysis framework
for the first phase of the experiment, but at the same time we are
planning for a 3D version of Commander that will be suitable for the
second phase of COMAP. While the intensity mapping field clearly is in
its infancy today, we do believe that it is only a matter of time
before we will see high signal-to-noise data coming out of these
experiments as well, and then the methods described will be
essential. Furthermore, from the \emph{Planck} experience, we know
that it takes about a decade to make such an approach work properly,
and so it is probably wise to get started.

%\vspace*{-0.7cm}

\end{document}